# A Fast Heuristic Algorithm Based on Verification and Elimination Methods for Maximum Clique Problem

Sabu .M Thampi [*]
*L.B.S College of Engineering,
Kasaragod, Kerala-671542, India*
smtlbs@yahoo.co.in

Murali Krishna P
*LBS College of Engineering
Kasaragod, Kerala-671542, India*
talktomkp@hotmail.com

**Abstract**

*A clique in an undirected graph G= (V, E) is a subset $V' \subseteq V$ of vertices, each pair of which is connected by an edge in E. The clique problem is an optimization problem of finding a clique of maximum size in graph. The clique problem is NP-Complete. We have succeeded in developing a fast algorithm for maximum clique problem by employing the method of verification and elimination. For a graph of size N there are $2^N$ sub graphs, which may be cliques and hence verifying all of them, will take a long time. Idea is to eliminate a major number of sub graphs, which cannot be cliques and verifying only the remaining sub graphs. This heuristic algorithm runs in polynomial time and executes successfully for several examples when applied to random graphs and DIMACS benchmark graphs.*
.

## 1. Introduction

A clique in an undirected graph G= (V, E) is a subset $V' \subseteq V$ of vertices, each pair of which is connected by an edge in E. The clique problem is an optimization problem of finding a clique of maximum size in graph. The Maximum Clique Problem (MCP) is a hard combinatorial problem, classified as NP-Complete. Most Theoretical Computer Scientists believe that NP-Complete problems are intractable. The reason is that if any single NP-Complete problem can be solved in polynomial time, every NP-Complete problem has a polynomial-time algorithm. The question whether NP = P? remains unanswered. It is obvious that $P \subset NP$, but whether $NP \subset P$? is unknown.

The maximum clique problem has many practical applications in science and engineering. These include project selection, classification, fault tolerance, coding, computer vision, economics, information retrieval, signal transmission, and alignment of DNA with protein sequences. A major application of the maximum clique problem occurs in the area of coding theory [1]. The goal here is to find the largest binary code, consisting of binary words, which can correct a certain number of errors.

A maximal clique is a clique that is not a proper set of any other clique. A maximum clique is a maximal clique that has the maximum cardinality. In many applications, the underlying problem can be formulated as a maximum clique problem while in others a subproblem of the solution procedure consists of finding a maximum clique. This necessitates the development of fast and approximate algorithms for the problem. Over the past decades, research on the maximum clique and related problems yielded many interesting and profound results. However, a great deal remains to be learned about the maximum clique problem.

The papers on finding cliques in a graph are a mixture of exact and heuristic methods. Carraghan and Pardalos [2] proposed a very simple and effective algorithm for the maximum clique problem. This algorithm was used as a benchmark in the Second DIMACS Implementation Challenge [3]. The algorithm initially searches the whole graph *G* considering the first vertex *v*1 and finds the largest clique *C*1 that contains *v*1. Then *v*1 is not further considered since it is not possible to find a larger clique containing *v*1. The algorithm next searches the graph considering *v*2 and finds *C*2, the largest clique in this subgraph that contains *v*2. The algorithm proceeds until no clique, larger than the present, can be found.

Balas and Niehaus [3] note that finding the maximum clique in the union of two cliques is solvable by matching techniques, and provide a heuristic based on this. Gibbons, Hearn, and Pardalos [3] look at a continuous-variable formulation of the problem and provide a heuristic based on rounding. Both Grossman and Jagota [4], Sanchis and Ganesan [5] look at neural network approaches. Soriano and Gendreau [6] look at a different type of general heuristic technique: tabu search. Goldberg and Rivenburgh [7] use a restricted backtracking method to provide a tradeoff between quality of clique and completeness of search. Homer and Peinado [8] develop a polynomial time heuristic based on subgraph exclusion and apply it to very large instances. Patric and Ostergard [9] present a branch-and-bound algorithm for the maximum clique problem. Some papers used different relaxations of the clique problem in order to create improved exact algorithms. Balas, Ceria, Cornuejols, and Pataki [10] use a "Lift and Project" method to find good bounds. Mannino and Sassano [11] use an edge projection method to create both a heuristic and an improved branch and bound exact algorithm. Bourjolly, Gill, Laporte and Mecure [12] look at strengthening relaxations of a quadratic 0-1 formulation. Brockington and Culberson [13] are concerned not with solving clique problems but with generating them. They show how to "hide" large cliques in graphs that look like random graphs.

---
[8]Corressponding Author

No polynomial-time algorithm has yet been discovered for an NP-Complete problem, nor has any one yet been able to prove a super polynomial-time lower bound for any of them. The main goal of this paper is to present an experimental study for solving the Maximum Clique Problem (MCP) using a *Verification and Elimination* method. If the algorithm gives an output then it will be the maximum size clique in that given graph, but we failed to provide an upper bound for the time it will take. Hence the algorithm is heuristic.

The remainder of this paper is organized as follows. Section 2 introduces the solution to the clique problem. Section 3 reports the computational results and discusses the experimental analysis. Finally, we draw conclusions in section 4.

## 2. Solution to Clique Problem

For a graph of size N there are $2^N$ subgraphs, which may be cliques and hence verifying all of them, will take a long time. Idea is to eliminate a major number of subgraphs, which cannot be cliques and verifying only the remaining subgraphs.

In a graph of size N, there are exactly $_NC_K$ subgraphs of size K.

$$\text{So total number of subgraphs in G} = \sum_{K=1}^{N} {}_NC_K = 2^N$$

This is exponential. So verifying all of the subgraphs will take a long time, because number of verifications required is not a polynomial in N. For decreasing the time required we can avoid a few verifications.

This algorithm uses the following important properties of clique.
- Every graph contains at least one clique.
- In a clique of size M, all the vertices have the degree M-1.
- If the maximum size of any vertex is M, there cannot be a clique of Size > M+1.
- If there is a clique of size K there are cliques of any Size < K in the same graph.
- Conversely, if there is no clique of Size K, there will not be a clique of Size > K.

### 2.1 Description of the Algorithm

The number of vertices in the clique is known as the size, N of the clique. There are three procedures known as FINDCLIQUE, SELECT and ISCLIQUE are used in the algorithm. First one calls the second algorithm and second in turn calls the third. The output will be the maximum clique.

**FINDCLIQUE** (N)
1. begin
2. m ← maximum degree of any vertex in G
3. lb ← 0
   ub ← m
   s ← nil
   clique ← nil
   iclique ← nil   /* iclique - intermediate clique*/
4. mid ← (lb + ub) / 2
5. k ← number of vertices which have degree >= mid
   s ← vertices with degree >= mid
6. if k > mid
   if **SELECT**(s, k, mid+1, iclique)
   cliquesize ← mid+1
   clique ← iclique
   lb ← mid+1
     else
            ub ← mid -1
     else
            ub ← mid -1
7. if lb <= ub
       mid ← (lb + ub) / 2
       goto step 5
8. print cliquesize
9. print clique
10. end
/*******************************/

**SELECT** (s, k, L, iclique)
1. begin
2. t ← first L vertices from s
/* first combination of vertices */

```
3. if ISCLIQUE (t, L)
      iclique ← t
      return true
   else
      t ← next combination from s
      if t= =nil
         return false
4. goto step 3
5. end
```

/*******************************/
**ISCLIQUE** (t, L)
1. begin
2. find a graph G' from G, which is
   induced by t
3. if degree of all the vertices in G'
   are equal to L-1
      return true
   else
      return false
4. end
/*******************************/

The different steps of the **FINDCLIQUE** algorithm are summarized as follows:

i. It finds the maximum degree m, sets lb to 0, ub to m and (lb + ub)/2 to mid
ii. Check whether there is a clique of size mid+1. if it is there, no need to verify the subgraphs of size <= mid and hence sets lb to mid+1
   If it is not there no need to consider the subgraphs with size > mid.
      Hence sets ub to mid-1.
iii. If lb<=ub, sets (lb + ub)/2 to mid and repeats the step ii.
iv. The clique, which is found just before when lb becomes > ub is the clique of maximum size in the given graph G.
v. Print the size, and vertices in that clique.

The checking of the existence of the clique of size mid +1 (step ii) is made by the **SELECT** algorithm. This algorithm finds the different combinations of the selected vertices and calls the algorithm **ISCLIQUE** for checking whether that combination of the vertices form a clique or not. If the degree of every vertex, in the induced sub graph by that combination is equal to mid, **ISCLIQUE** returns TRUE.

Consider the graph shown in figure 1. In this graph N = 16 and the vertices are numbered as 1, 2..., 16.

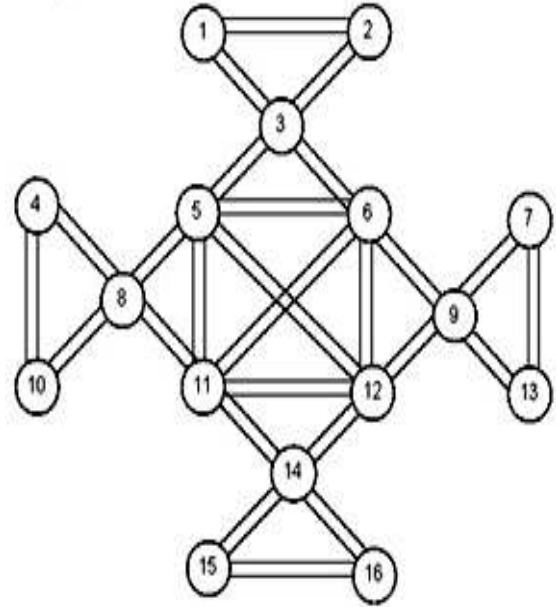

**Figure 1: Example Graph**

1. Maximum degree m = 5, lb=0, ub= 5
2. Number of vertices with degree>=5 is 4 which is < 6. Hence there cannot be a clique of size 6. Set mid= (0+5)/2=2.
3. Number of vertices with degree >=2 is 16. So there may be clique of size 3. Select first 3 vertices. Vertices 1, 2 & 3 form a clique. Hence set lb to 3.
4. mid= (3+5)/2=4
5. Number of vertices with degree >= 4 is 8. They are 3, 5, 6, 8, 9, 11, 12, and 14. None of their combination forms a clique of size 5. Hence set ub to 3
6. lb<=ub, therefore mid = (3+3)/2=3.
7. Number of vertices with degree >= 3 is 8. Out of these 8 vertices the combination 5, 6, 11 and 12 forms a clique. Hence set lb to 4.
8. Now lb=4 and ub = 3
9. Since lb >ub, the last found clique is the clique of maximum size. Therefore, {5, 6, 11, 12} is the solution.

### 2.2 Time Complexity of the Algorithm

The algorithm FINDCLIQUE works similar to that of *binary search*. For maximum degree M, the time complexity of the algorithm FINDCLIQUE is $O(\log_2(M))$, which is polynomial. If the size of the subgraph to be verified is k, the time complexity of the algorithm ISCLIQUE is $O(k^2)$, which is again polynomial. The algorithm SELECT is based on combinations. In worst case, all the $_NC_K$ combinations may have to be verified. But whenever a clique is found, the algorithm will *return,* neglecting the remaining combinations. So, average time complexity of this algorithm is practically polynomial.

## 3. Results and Discussion

To properly evaluate the new heuristic algorithm, it is necessary to carry out practical experiments. We will here show how the new algorithm performs for random graphs and DIMACS benchmark graphs. The new algorithm is compared with the algorithm in [9] in the performance tests. The computational results with random graphs are presented in Table 1. The new algorithm is implemented in C language. The entries in Table 1 show the average run time in CPU seconds on a Pentium PC with 256MB memory and Linux Operating System. For each entry, 8 random graphs were constructed and used as input. The graphs were constructed by considering each pair of vertices and the edge density.

Looking at the results in Table 1, the new algorithm is clearly the fastest for the random graphs. When the edge density increases the speed of the algorithm also gets increased.

We extensively tested the new heuristic algorithm on DIMACS benchmark graphs to show its efficiency. The benchmark graphs were downloaded from URL:ftp://dimacs.rutgers.edu/pub/challenge/graph/benchmarks/clique/. Table 2 compares the performance of new algorithm and the algorithm described in [9] for the DIMACS graphs. The outputs generated by the program for a few DIMACS Benchmark graphs are listed in Table 3.

### Table 1: Random Graphs

| Vertices n | Edge Density d | [9] | New |
|---|---|---|---|
| 100 | 0.6 | 0.01 | 0.01 |
| 100 | 0.7 | 0.03 | 0.02 |
| 200 | 0.4 | 0.03 | 0.025 |
| 200 | 0.5 | 0.09 | 0.079 |
| 300 | 0.3 | 0.05 | 0.04 |
| 300 | 0.4 | 0.21 | 0.205 |
| 500 | 0.2 | 0.09 | 0.08 |
| 500 | 0.3 | 0.37 | 0.362 |

## 4. Conclusion

We have in this paper introduced a heuristic algorithm for the maximum clique problem based on verification and elimination method. We have found our new heuristic to perform competitively when compared to other algorithms on the DIMACS benchmark graphs and random graphs. Apart from this, there certainly is room for further research and enhancements. The task of selecting combinations and verifying vertices can be parallelized, and if developed a parallel algorithm for that, it will surely run faster than new algorithm.

### Table 2: DIMACS Graphs

| DIMACS Graph | n | d | Size | [9] | New |
|---|---|---|---|---|---|
| brock200_1 | 200 | 0.75 | 21 | 54.29 | 31.38 |
| brock200_2 | 200 | 0.50 | 12 | 0.05 | 0.03 |
| c_fat200_1 | 200 | 0.08 | 12 | 0.01 | 0.005 |
| c_fat500_1 | 500 | 0.04 | 14 | 0.07 | 0.06 |
| hammings6_2 | 64 | 0.90 | 32 | 0.01 | 0.01 |
| hammings6_4 | 64 | 0.35 | 4 | 0.01 | 0.013 |
| keller4 | 171 | 0.65 | 11 | 0.50 | 0.46 |
| p_hat300_1 | 300 | 0.24 | 8 | 0.04 | 0.036 |
| p_hat500_1 | 500 | 0.25 | 9 | 0.29 | 0.21 |
| san200_0.7_1 | 200 | 0.70 | 30 | 0.56 | 0.46 |
| san400_0.5_1 | 400 | 0.50 | 13 | 0.03 | 0.02 |
| MANN_a9 | 45 | 0.93 | 16 | 0.01 | 0.01 |
| johnson8_2_4 | 28 | 0.56 | 4 | 0.01 | 0.01 |
| johnson8_4_4 | 70 | 0.77 | 14 | 0.01 | 0.01 |

### Table 3: Output for DIMACS Graphs

| DIMACS Graph | Output Generated |
|---|---|
| brock200_1 | The size of the maximum clique: 21.<br>The nodes of this clique are:<br>4 26 32 41 46 48 83 100 103 104 107 120 122 132<br>137 138 144 175 180 191 199 |
| c_fat200_1 | The size of the maximum clique: 12.<br>The nodes of this clique are:<br>6 7 43 44 80 81 117 118 154 155 191 192 |
| san400_0.5_1 | The size of the maximum clique: 13.<br>The nodes of this clique are:<br>12 40 97 118 194 199 222 229 254 261 327 349 395 |
| MANN_a9 | The size of the maximum clique: 16.<br>The nodes of this clique are:<br>3 4 6 8 11 14 16 19 22 27 28 33 34 39 41 45 |
| keller4 | The size of the maximum clique: 11.<br>The nodes of this clique are:<br>27 37 41 53 62 66 135 140 150 156 168 |